\documentclass[twocolumn,preprintnumbers]{revtex4}
\usepackage{graphicx}% Include figure files
\usepackage{dcolumn}% Align table columns on decimal point
\usepackage{bm}% bold math

\begin{document}

\preprint{}

\title{Kagom\'e in triangular lattice: \\
electronic state of CoO$_2$ layer with hexagonal structure}% Force line breaks with \\

\author{W. Koshibae}
%\email{wataru@imr.tohoku.ac.jp}
\author{S. Maekawa}
\affiliation{%
Institute for Materials Research,
Tohoku University, Sendai 980-8577, Japan
}%

\date{}
\begin{abstract}
The electronic state in layered cobalt oxides with hexagonal structure is examined.  
We find that the electronic structure reflects the nature of the Kagom\'e lattice 
hidden in the CoO$_2$ layer which consists of stacked triangular lattices of oxygen ions 
and of cobalt ions.  A fundamental model for the electron system is proposed 
and the mechanism of the unique transport and magnetic properties of the cobalt oxides are discussed 
in the light of the model.  
\end{abstract}

\maketitle

Cobalt oxides with layered hexagonal structure 
have attracted much attention.  The oxides 
Na$_x$CoO$_2$, [Bi$_{2-x}$Pb$_x$Sr$_2$]$_y$CoO$_2$ and [CaCo$_3$]$_x$CoO$_2$ 
exhibit large thermopower and have potentials for thermoelectric materials~\cite{terasaki,yama,masset}. 
In addition, the anomalous high-temperature Hall effect has been observed in 
Na$_{0.68}$CoO$_2$~\cite{wang2}.  
It is known that Na$_{0.75}$CoO$_2$ shows 
a magnetic transition at 22 K~\cite{motohashi}, and 
[Bi$_{2-x}$Pb$_x$Sr$_2$]$_y$CoO$_2$ is ferromagnetic below $T_C=$ 3.2K~\cite{tsukada}.  
Recently, superconductivity~\cite{takada} has been discovered 
in water-intercalated Na$_{0.35}$CoO$_2\cdot$1.3H$_2$O.  
Since then, experimental and theoretical studies have been done 
extensively~\cite{wang1,wang2,cava,sato,yoshimura,singh2,hu,shastry1,baskaran,ogata,lee}.
Several authors~\cite{shastry1,baskaran,ogata,lee} have examined 
the electronic state based on a single band model in triangular lattice of Co ions 
of the CoO$_2$ layer.  

In this paper we show that a Kagom\'e lattice structure is hidden in the CoO$_2$ layer, 
and the electronic state is based on the Kagom\'e lattice but not the single band model 
in the triangular lattice.  
We propose a fundamental model for the electron system and discuss the mechanism 
of the unique transport and magnetic properties in the cobalt oxides.  
 
\begin{figure}
\includegraphics[width=8.0cm]{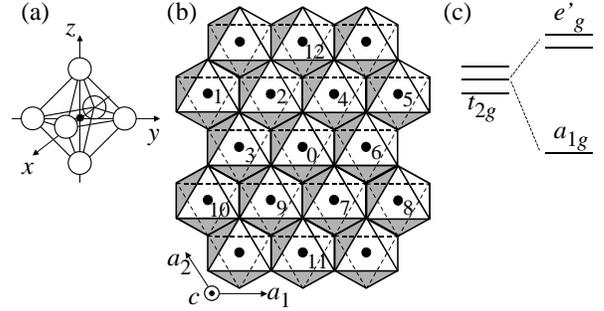}
\vspace{-10pt}
\caption{\label{fig1}
(a) CoO$_6$ octahedron.  Solid and open circles indicate 
cobalt and oxygen ions, respectively.  
(b) CoO$_2$ layer.  
 $c$ and $a_1$ axes are along (1,1,1) and ($-$1,1,0) directions in 
$xyz$ coordinate system shown in (a).  
The numbers (0 $\sim$ 12) on solid circles are the labels of  
Co sites.  (c) The crystal-field 
splitting of the distorted CoO$_6$ octahedron.  
$e'_g$ is used to distinguish the $e_g$ (${x}^2-{y}^2$ and $3{z}^2-{r}^2$) states.  
}
\end{figure}

The crystal structure is presented in Fig.~\ref{fig1}.  
The CoO$_2$ layer is formed by the edge-shared CoO$_6$ octahedra 
which are compressed along $c$-axis.  
The rhombohedral distortion of the CoO$_6$ octahedra are estimated 
by the deviation of the O-Co-O bond angle from 90$^\circ$, 95$^\circ\sim 99^\circ$~\cite{singh,miyazaki,kajitani,ono,yama}.  
The distortion leads to the crystal-field splitting in $t_{2g}$ states of 3$d$ electrons 
as shown in Fig.~\ref{fig1}(c).  The wave functions are expressed as 
\begin{equation}
(|xy> + |yz> + |zx>)/ \sqrt{3} 
\label{a1g}
\end{equation}
for the $a_{1g}$ state and 
\begin{equation}
\label{e'g}
(|xy> + e^{\pm i{2\pi\over 3}}|yz> + e^{\pm i{4\pi\over 3}}|zx>)/ \sqrt{3},
\end{equation}  
for the doubly degenerate $e'_g$ states 
where $|xy>$, $|yz>$ and $|zx>$ denote the wave functions of 
the $t_{2g}$ states.  The $a_{1g}$ state extends to 
the $c$-axis whereas 
the $e'_g$ states spread over the plane perpendicular to the $c$-axis.  
Since the apex oxygens approach the plane 
in the distorted CoO$_6$ octahedra, 
the $a_{1g}$ state is stabilized~\cite{foot} for an electron.

%%%%%%%%%%%%%%%%%%%%%%%%%%%%%%%%%%%%%%%%%%%%%%%%%%%%%%%%%%%%%%
%%%%%%%%%%%%%%%%%%%%%%%%%%%%%%%%%%%%%%%%%%%%%%%%%%%%%%%%%%%%%%

The band calculation~\cite{singh} in Na$_{0.5}$CoO$_2$ 
has shown that the energy splitting between $a_{1g}$ and $e'_g$ states at the $\Gamma$ point is $\sim$ 1.6eV which is the total band width of the $t_{2g}$ manifold and the $a_{1g}$ state is higher than the $e'_g$ states.  
This fact shows that the energy splitting does not originate in 
the crystal field due to the distortion but is determined by the kinetic energy 
of electrons.  

\begin{figure}[t]
\includegraphics[width=6.5cm]{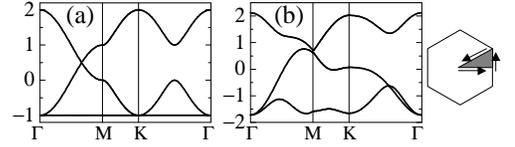}
\vspace{-10pt}
\caption{\label{fig2}
(a) Dispersion relation of Eq.~(\ref{dp}) for $t=1$.  
(b) Dispersion relation of $E_{\vec{k},\gamma,\gamma'}$ for $t=1$, $t_{dd}=-1$, 
$t_1=-0.035$ and $t_2/t_1=-2.5$.  
The symbols $\Gamma$, M and K denote the $k$-points, 
(0,0), $(\pi,0)$ and $(\pi,\pi/\sqrt{3})$ in Cartesian coordinates.  
}
\end{figure}

Let us consider the hopping-matrix-elements between neighboring 
3$d$ orbitals of cobalt ions neglecting the rhombohedral distortion.  
There are two mechanisms for the hopping of an electron: 
one is the hopping integral between adjacent 3$d$ orbitals, and another is 
owing to the hopping between a 3$d$ orbital of a cobalt ion and a 2$p$ one 
of an oxygen ion.  
First, let us consider the latter mechanism,  
i.e., the hopping of a 3$d$ electron through the 2$p$ orbital on the neighboring oxygen.  
The CoO$_2$ layer in the hexagonal structure is expressed as a 
triangular lattice of cobalt ions sandwiched by those of oxygen ions, 
i.e., both the upper and lower layers of oxygens form triangular lattices.  
The lower layer is drawn by broken lines in Fig.~1(b).  
In the following, the $t_{2g}$ orbitals on the $i$-th cobalt ions are 
expressed as $|xy,i>$, $|yz,i>$ and $|zx,i>$, respectively.  
The state $|zx,0>$ has a hopping matrix element with 
$|yz,6>$ through the 2$p_z$ orbital of the oxygen ion which exists 
in the lower layer and shares CoO$_6$ octahedra involving cobalt ions 
0 and 6, respectively.  
The hopping matrix element ($t$) is expressed as 
$t \sim \left(t_{pd}\right)^2/\Delta\ (>0)$,  where 
$\Delta$ is the energy level of the 2$p_z$ orbital measured from that of $t_{2g}$ states 
and $t_{pd}$ is the hopping integral between the 2$p_z$ and $|zx,0>$ (or $|yz,6>$) orbitals.  
There also exists a hopping matrix element between 
$|yz,0>$ and $|zx,6>$ which is due to the 2$p_z$ orbital of an oxygen ion on the upper 
layer.  On the other hand, there is no hopping matrix element between $|xy,0>$ and 
$|xy,6>$ because of the symmetry.  
In the same way, we find the hopping matrix elements between the following pairs of orbitals:
$(|xy,0>,|zx,2>)$, 
$(|zx,0>,|xy,2>)$, 
$(|xy,0>,|yz,4>)$, and 
$(|yz,0>,|xy,4>)$.  
As a result, the hopping matrices of a 3$d$ electron in $\vec{u}$, $\vec{v}$ and $\vec{u}+\vec{v}$ 
directions where $\vec{u}$ and $\vec{v}$ are the elementary translation vectors of the cobalt 
triangular lattice, are expressed as  
\begin{eqnarray} 
\begin{array}{cc}
&
\begin{array}{ccc}
xy&yz&zx\\
\end{array}
\\
\begin{array}{c}
xy\\
yz\\
zx\\
\end{array}
&
\left(
\begin{array}{ccc}
0&0&0\\
0&0&t\\
0&t&0\\
\end{array}
\right),\\
\end{array}
\begin{array}{c}
\begin{array}{ccc}
xy&yz&zx\\
\end{array}
\\
\left(
\begin{array}{ccc}
0&0&t\\
0&0&0\\
t&0&0\\
\end{array}
\right)\\
\end{array}
\rm{and}
\begin{array}{c}
\begin{array}{ccc}
xy&yz&zx\\
\end{array}
\\
\left(
\begin{array}{ccc}
0&t&0\\
t&0&0\\
0&0&0\\
\end{array}
\right)\\
\end{array},
\label{tdp} 
\end{eqnarray} 
respectively.  
In the Fourier-transformed representation, the tight binding Hamiltonian is written as 
\begin{equation} 
H_t = \sum_{\vec{k},\sigma,\gamma\gamma'}
\varepsilon_{\vec{k}\gamma\gamma'}c_{\vec{k}\sigma\gamma}^\dagger c_{\vec{k}\sigma\gamma'},
\label{dp} 
\end{equation} 
with
\begin{eqnarray} 
\varepsilon_{\vec{k}}=2t\left[
\begin{array}{ccc}
0&\cos(k_1+k_2) &\cos k_2 \\
\cos(k_1+k_2)&0&\cos k_1 \\
\cos k_2&\cos k_1 &0 \\
\end{array}
\right],
\label{dpq} 
\end{eqnarray} 
where $k_1$, and $k_2$ are the component of the wave vector $\vec{k}$ in the representation of the reciprocal lattice of the triangular lattice spanned by $\vec{u}$ and $\vec{v}$.  
$\gamma$, $\gamma'$ (= $xy$, $yz$, $zx$) and $\sigma$ (= $\uparrow$, $\downarrow$) 
are the indices for the $t_{2g}$ orbitals and electron spin, respectively, and 
$c^\dagger_{\vec{k}\sigma\gamma}$ ($c_{\vec{k}\sigma\gamma'}$) denotes 
creation (annihilation) operator of an electron with $\vec{k}$, $\sigma$ and $\gamma$ 
($\gamma'$).  
Eq.~(\ref{dpq}) shows the well-known dispersion relation of the Kagom\'e lattice (see Fig.~\ref{fig2}).   
This means that the Kagom\'e lattice structure stays in hiding in the triangular lattice of cobalt ions.  
Let us discuss how to realize the Kagom\'e lattice in 
the motion of an electron given by Eq.~(\ref{tdp}).  
An electron in the state $|zx,1>$ can go to $|yz,2>$ and $|xy,3>$.  
An electron in $|yz,2>$ can go to $|xy,12>$ and $|xy,3>$ 
but cannot go to any orbitals on cobalt 0 due to the symmetry.  
In this way, an electron starting from $|zx,1>$ 
propagates through the $t_{2g}$ orbitals on cobalt ions, 
1 $\sim$ 12, using Eq.~(\ref{tdp}), 
and thus the trace of the motion forms a Kagom\'e lattice (see Fig.~3(a)).   
The triangle made of the states $|zx,1>$, $|yz,2>$ and $|xy,3>$ is an elementary unit 
of the Kagom\'e lattice.    
Therefore, the energy scheme of the triangle determines that at (0,0) in the $k$ space.  
The eigenstates of the triangle are 
\begin{equation}
(|xy,3> + |yz,2> + |zx,1>)/ \sqrt{3} 
\label{a1gt}
\end{equation}
with the eigenvalue 2$t$ and 
\begin{equation}
\label{e'gt}
(|xy,3> + e^{\pm i{2\pi\over 3}}|yz,2> + e^{\pm i{4\pi\over 3}}|zx,1>)/ \sqrt{3}, 
\end{equation}  
with $-t$.  
The eigenstates have the $a_{1g}$ and $e'_g$ symmetries, respectively.  
Note that they are completely different from the states Eqs.~(\ref{a1g}) and (\ref{e'g}).  
The eigenstates in Eqs.~(\ref{a1gt}) and (\ref{e'gt}) lie on the top and bottom of the 
band, respectively. This is a character of the Kagom\'e lattice structure.

%%%%%%%%%%%%%%%%%%%%%%%%%%%%%%%%%%%%%%%%%%%%%%%%%%%%%%%%%%%%%%
%%%%%%%%%%%%%%%%%%%%%%%%%%%%%%%%%%%%%%%%%%%%%%%%%%%%%%%%%%%%%%

\begin{figure}[t]
\includegraphics[width=5.5cm]{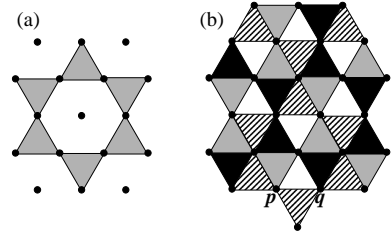}
\vspace{-10pt}
\caption{\label{fig3}
Kagom\'e lattice in the triangular lattice of cobalt ions.  
(a) Solid circles indicate the cobalt ions in Fig.~1(b).  
Gray triangles form a Kagom\'e lattice which is made by a trace of the travel of an electron 
starting from $|zx,1>$ (see text).  
(b) Layout of the four (gray, black, hatched and white) Kagom\'e lattices.  
}
\end{figure}

When an electron propagates starting from $|zx,1>$, the trace forms another Kagom\'e lattice 
which is drawn by black triangles in Fig.~\ref{fig3}(b).  
Following the procedure, we obtain four Kagom\'e lattices in the triangular lattice of cobalt ions 
as shown in Fig.~\ref{fig3}(b).  Because the unit cell of the Kagom\'e lattice is four times as large as 
that of the triangular lattice of cobalt ions, the four Kagom\'e lattices complete the Hilbert space.

%%%%%%%%%%%%%%%%%%%%%%%%%%%%%%%%%%%%%%%%%%%%%%%%%%%%%%%%%%%%%%
%%%%%%%%%%%%%%%%%%%%%%%%%%%%%%%%%%%%%%%%%%%%%%%%%%%%%%%%%%%%%%

In the CoO$_2$ layer where the CoO$_6$ octahedra share the edges as shown in Fig.~\ref{fig1}, 
the hopping integral between adjacent $3d$ orbitals should be taken into account 
to analyze the band structure as well.  
Between the cobalt ions 1 and 2, the leading term of the hopping matrix element ($t_{dd}$) occurs 
between  $|xy,1>$ and $|xy,2>$.  
The sign of the hopping matrix element $t_{dd}$ is negative due to the configuration of the orbitals 
on the hexagonal CoO$_2$ layer.  
Although there exist the other hopping matrix elements between the ions 1 and 2, 
e.g., between $|zx,1>$ and $|yz,2>$, 
their magnitude may be much smaller than the leading term.  
The hopping between $xy$ orbitals forms a one-dimensional chain along $a_1$-axis.  
In the same way, the hopping between $yz$ ($zx$) orbitals 
forms another chain along $a_2$-axis (the direction of $\vec{u}+\vec{v}$).  
Consequently, the hopping matrix in the Fourier-transformed expression 
is diagonal and is written as 
$2t_{dd}\cos(k_1)$ for ($xy,xy$), $2t_{dd}\cos(k_2)$ for ($yz,yz$) and 
$2t_{dd}\cos(k_1-k_2)$ for ($zx,zx$) component, respectively.  
Note that the hopping matrix does not give the energy-level splitting at (0,0) in the $k$ space.  

\begin{figure}[t]
\includegraphics[width=3.0cm]{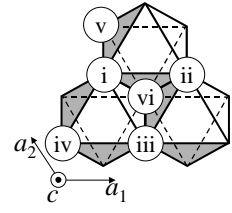}
\vspace{-10pt}
\caption{\label{fig4}
Triangular lattices of oxygen ions in CoO$_2$ layer.  
Cobalt ions are not drawn.  i $\sim$ iv 
(v $\sim$ vi) are on the 
upper (lower) triangular lattice of oxygen ions.
}
\end{figure}

For more detailed analysis, we introduce the effect of the hopping integral of 2$p$ orbitals between 
neighboring oxygen ions, which depends on the configuration of adjacent 2$p$ orbitals.  
Let us consider the configuration of 2$p_z$ orbitals on the oxygen ions labeled 
i $\sim$ vi in Fig.~\ref{fig4} where the relation between $xyz$ and $a_1a_2c$ coordinate systems 
are the same as that in Fig.~\ref{fig1}.  
The 2$p_z$ orbitals on i, ii and v are parallel to each other but those on i and iii are not.  
Using the table by Slater and Koster~\cite{sk}, we find two kinds of hopping integrals as follows:
$
t_{pp,1}=V_{pp\pi}
$
for the neighboring pair of oxygen ions, (i,ii) and (i,v), and
$
t_{pp,2}=(1/2)\left(V_{pp\sigma}+V_{pp\pi}\right)
$
for (i,iii), (i,iv) and (i,vi), where 
$V_{pp\sigma}/V_{pp\pi}=-4$, $t_{pp,2}/t_{pp,1}=-2.5$ and the sign of $t_{pp,1}$ is negative.  
Following the procedure, all of the hopping integrals of 2$p$ orbitals 
between neighboring oxygen ions are expressed as $t_{pp,1}$ and $t_{pp,2}$, which 
lead to the hopping matrix elements, $t_1$ and $t_2$,  
of the 3$d$ electron to the second nearest neighbors.  
The hopping matrix element is derived in the form 
$t_n \sim (t_{pd})^2t_{pp,n}/(\Delta)^2$ where $n$ = 1 or 2.  
As a result, we obtain the hopping matrix $E_{\vec{k},\gamma,\gamma'}$ of the CoO$_2$ layer:  
\begin{eqnarray*} 
E_{\vec{k},\gamma,\gamma}&=&
2\left(t_{dd}+2t_2\right)\cos k^{\gamma,\gamma}_a\\
&&+2(t_1+2t_2)\left[\cos k^{\gamma,\gamma}_b
+\cos\left(k^{\gamma,\gamma}_a+k^{\gamma,\gamma}_b\right)\right]\\
&&-2t_2\left[\cos\left(2k^{\gamma,\gamma}_a+k^{\gamma,\gamma}_b\right)
+\cos\left(k^{\gamma,\gamma}_a-k^{\gamma,\gamma}_b\right)\right]\\
E_{\vec{k},\gamma,\gamma'}&=&
2t\cos k^{\gamma,\gamma'}_b+2t_1\cos 2k^{\gamma,\gamma'}_b\\
&&+2t_2\left[\cos\left(k^{\gamma,\gamma'}_a+2k^{\gamma,\gamma'}_b\right)
+\cos\left(k^{\gamma,\gamma'}_a-k^{\gamma,\gamma'}_b\right)\right],
\label{hamq} 
\end{eqnarray*} 
where 
$k^{xy,xy}_a=k^{xy,zx}_a=k_1$, $k^{xy,xy}_b=k^{xy,zx}_b=k_2$, 
$k^{yz,yz}_a=k^{xy,yz}_a=k_2$, $k^{yz,yz}_b=k^{xy,yz}_b=-(k_1+k_2)$, 
$k^{zx,zx}_a=k^{yz,zx}_a=-(k_1+k_2)$ and $k^{zx,zx}_b=k^{yz,zx}_b=k_1$, respectively.  
The dispersion relation reproduces 
the band structure calculated by Singh~\cite{singh}.  
Although the more parameters may result in the more quantitative agreement 
with the band structure, it is not the purpose of this paper.  

The dispersion relation Fig.~\ref{fig2}(b) clearly shows that 
the upper lying band takes over 
the nature of the Kagom\'e lattice structure 
hidden in the triangular lattice of cobalt ions (see Fig.~\ref{fig3}) 
despite of the presence of $t_{dd}$ and 
the correction of the higher order terms $t_1$ and $t_2$.   
Therefore, it is of crucial importance to study the effect of the Kagom\'e lattice structure 
to clarify the electronic state in the CoO$_2$ layer.  

%%%%%%%%%%%%%%%%%%%%%%%%%%%%%%%%%%%%%%%%%%%%%%%%%%%%%%%%%%%%%%
%%%%%%%%%%%%%%%%%%%%%%%%%%%%%%%%%%%%%%%%%%%%%%%%%%%%%%%%%%%%%%

So far, we have not discussed the effect of Coulomb interaction.  
In reality, the effective on-site Hubbard $U$ of 5-8 eV is the largest 
scale of energy in the electron system.  
Let us discuss the effect of the strong Coulomb interaction in the Kagom\'e lattice structure 
shown in Fig.~\ref{fig3}.  
The four Kagom\'e lattices share the edges each other.  
Each edge consists of two cobalt ions on both ends, 
e.g., the cobalt ions $p$ and $q$ form the edge shared by white and hatched Kagom\'e lattices.  
The 3$d$ orbitals $|yz,p>$ and $|zx,q>$ belong to the white Kagom\'e lattice.  
On the other hand, $|zx,p>$ and $|yz,q>$ are in the hatched Kagom\'e lattice.  
The orbitals on a cobalt ion belonging to the different Kagom\'e lattices are orthogonal to each other.  
The on-site Coulomb interaction brings about the following interaction between $p$ and $q$ when an electron is in the white Kagom\'e lattice and 
another is in the hatched one:
\begin{eqnarray*} 
\left(\vec{T}_p\cdot\vec{T}_q-{1\over 4}\right)
\left[J\left(\vec{S}_p\cdot\vec{S}_q+{3\over 4}\right)+ J'
\left(\vec{S}_p\cdot\vec{S}_q-{1\over 4}\right)\right],
\end{eqnarray*} 
where $\vec{S}_p$ ($\vec{S}_q$) is the electron spin on $p$ ($q$).  
The orbital state on $p$ ($q$) is described by the pseudo-spin 
operator $\vec{T}_p$ ($\vec{T}_q$) with 1/2 in magnitude,  
i.e., $|yz,p>$ ($|zx,p>$) 
is the eigenstate of $T^z_p$ with the eigenvalue $1/2$ ($-1/2$) 
and $|zx,q>$ ($|yz,q>$) 
is the eigenstate of $T^z_q$ with the eigenvalue $1/2$ ($-1/2$).  
$J$ and $J'$ are expressed as
$J=4t^2/(U'-K)$ and $J'=4t^2/(U'+K)$ with the Coulomb interaction $U'$ 
of the inter-orbitals and Hund's rule coupling $K$ of the $t_{2g}$ orbital on 
the cobalt ions.  
Due to the Hund's rule coupling, $J>J'$, i.e., there exists 
a ferromagnetic spin coupling with a singlet state of orbitals 
on the edge shared by the Kagom\'e lattices.  

We propose a fundamental model to study the electronic structure of 
the CoO$_2$ layer under the local constraint; $
\sum_{\sigma\gamma}c^\dagger_{i\sigma\gamma}c_{i\sigma\gamma}\geq 5.$
The Hamiltonian is expressed as:
\begin{equation}
H=H_{t}+H_{J},
\end{equation}
with  
\begin{eqnarray*}
H_{J}=\sum_{nm}\left(
H^{(1)}_{2n\vec{u}+m\vec{v}}+
H^{(2)}_{m\vec{u}+2n\vec{v}}+
H^{(3)}_{2n(\vec{u}+\vec{v})+m\vec{v}}
\right),
\end{eqnarray*} 
where $n$ and $m$ are integer numbers and 
\begin{eqnarray*}
H^{(I)}_i=J\sum_{\vec{\delta}(I)}
&&\left(\vec{T}^{(I)}_i\cdot\vec{T}^{(I)}_{i+\vec{\delta}(I)}
-{1\over 4}n^{(I)}_in^{(I)}_{i+\vec{\delta}(I)}\right)\\ 
&&\times\left(\vec{S}_i\cdot\vec{S}_{i+\vec{\delta}(I)}+{3\over 4}n^{(I)}_in^{(I)}_{i+\vec{\delta}(I)}\right),
\end{eqnarray*} 
where $I$ is an index and $\vec{\delta}(1)=\pm\vec{u}$, 
$\vec{\delta}(2)=\pm\vec{v}$, and $\vec{\delta}(3)=\pm(\vec{u}+\vec{v})$ in the summation.  
The orbitals corresponding to the eigenstates of $T^{(I),z}_i$ are summarized in the Table~\ref{table1},  
and $n^{(I)}_i=n^{(I)}_{i+}+n^{(I)}_{i-}$ where 
$n^{(I)}_{i\pm}$ is the electron number in the eigenstates of $T^{(I),z}_i$.  

\begin{table}
\caption{
Relation between $t_{2g}$ ($xy$, $yz$ and $zx$) orbitals and eigenstates of $T^{(I),z}_i$    
with $i=n\vec{u}+m\vec{v}$.  
The letters with (without) the bracket denote the 
orbitals corresponding to the eigenstates in the case that $n+m$ is odd (even).  
}
\begin{tabular}{ddd}
\hline
$eigenvalue$  & 1/2  & -1/2 \\
\hline
I=1 & yz\ (zx) & zx\ (xy)  \\
I=2 & zx\ (xy) & xy\ (zx)  \\
I=3 & xy\ (yz) & yz\ (xy)  \\
\hline
\end{tabular}
\label{table1}
\end{table}

This model involves the ingredients for the unique transport and magnetic properties 
of the CoO$_2$ layer:
A spin-triplet with orbital-singlet  
is stabilized on a neighboring cobalt bond.  
The pairing mechanism acts in a different way from the so-called 
resonating-valence-bond picture 
discussed by several authors~\cite{shastry1,baskaran,ogata,lee},   
where the key is the singlet state of orbitals.  
Khaliullin and Maekawa~\cite{km} discussed 
a liquid state of $t_{2g}$ orbitals in a perovskite titanate.  
In the CoO$_2$ layer, 
the resonance and dynamics of the singlet states are developed in the Kagom\'e lattice 
but not in the triangular lattice.  
Note that the orbitals are characterized by four flavors, 
i.e., the four Kagom\'e lattices as shown in Fig.~\ref{fig3}(b), 
rather than three $t_{2g}$ states.  
Thus, the carrier doping may cause 
the superconductivity with spin-triplet~\cite{yoshimura}.  
This is based on the Kagom\'e lattice structure and  
is different from that on a single band model in 
a triangular lattice.  

The Kagom\'e lattice involves a triangle as a basic unit.  On the triangle, 
the mechanism by Kumar and Shastry~\cite{shastry1} 
for the anomalous Hall effect will be available and 
explains the experimtnts by Wang {\it et al.}~\cite{wang2}  

The triangle gives the $a_{1g}$ state Eq.~(\ref{a1gt}) as 
the upper lying band in the $t_{2g}$ manifold in the CoO$_2$ layer.  
However, a cobalt ion in the triangle is shared by three Kagom\'e lattices (see Fig.~\ref{fig3}).  
Consequently, the three $t_{2g}$ orbitals are identical.  
The orbital degree of freedom 
causes the large thermopower~\cite{koshibae1,koshibae2} at high temperatures.  

The Kagom\'e lattice structure clearly explains the non-symmetric nature of the band structure of 
the CoO$_2$ layer.   
When the effect of the Kagom\'e lattice becomes dominant, the bottom band, i.e., the flat band as shown in 
Fig,~\ref{fig2}(a) will play a crucial role on the electronic state.  
Mielke~\cite{mielke} have shown 
that the flat band with the Coulomb interaction has the ferromagnetic ground state 
at around half filling.  
A prospective system for the ferromagnet will be $d^1$ transition metal oxides, 
i.e., the layered titanates with iso-structure of the cobalt oxides.  
Although there exist many effects which disturb the flat band structure in reality, 
exploring of the ferromagnet based on the mechanism will be a challenging problem. 

In summary,  
we have shown that the Kagom\'e lattice structure is hidden in the CoO$_2$ layer.   
The electronic structure strongly reflects the nature of the Kagom\'e lattice but 
not the single band model with a triangular lattice.  
We have proposed a fundamental model for the electron system and discussed the mechanism 
of the unique transport and magnetic properties of the cobalt oxides.  

The authors are grateful to K. Tsutsui and T. Tohyama for useful discussions.  
This work was supported by Priority-Areas Grants from the Ministry 
of Education, Science, Culture and Sport of Japan and CREST.


\begin{thebibliography}{}
\bibitem{terasaki} 
I. Terasaki {\it et al.}, 
Phys. Rev. B{\bf 56},  R12685 (1997).
\bibitem{yama}        
T. Yamamoto {\it et al.},
Jpn. J. Appl. Phys. {\bf 39}, L747 (2000);
T. Yamamoto, Ph.D. thesis, University of Tokyo, 2001.
\bibitem{masset}
A. C. Masset {\it et al.},
Phys. Rev. B {\bf 62}, 166 (2000).
\bibitem{wang2} 
Y. Wang {\it et al.},
cond-mat/0305455.
\bibitem{motohashi} 
T. Motohashi {\it et al.}, 
Phys. Rev. B{\bf 67}, 64406 (2003).
\bibitem{tsukada}
I. Tsukada {\it et al.}, 
J. Phys. Soc. Jpn. {\bf 70}, 834 (2001).
\bibitem{takada}
K. Takada {\it et al.},
Nature {\bf 422}, 53 (2003).
\bibitem{wang1} 
Y. Wang {\it et al.},
Nature {\bf 423}, 425 (2003).
\bibitem{cava} 
R.E. Schaak {\it et al.},
cond-mat/0305450.
\bibitem{sato}
Y. Kobayashi, M. Yokoi and M. Sato, 
cond-mat/0305649, cond-mat/0306264.
\bibitem{yoshimura}
T. Waki {\it et al.},
cond-mat/0306036.
\bibitem{hu}
A. Tanaka and X. Hu, cond-mat/0304409.
\bibitem{singh2}
D. J. Singh, cond-mat/0304532.
\bibitem{shastry1}
B. Kumar and B. S. Shastry, cond-mat/0304210.
\bibitem{shastry2}
B. S. Shastry and B. I. Shraiman, Phys. Rev. Lett. {\bf 70}, 2004 (1993).  
\bibitem{baskaran}
G.Baskaran, cond-mat/0303649, cond-mat/0306569.
\bibitem{ogata}
M. Ogata, cond-mat/0304405.
\bibitem{lee}
Q.-H. Wang {\it et al.}, 
cond-mat/0304377.
\bibitem{singh}
D.J. Singh, Phys. Rev. B {\bf 61}, 13397 (2000).
\bibitem{miyazaki}
Y. Miyazaki {\it et al.},
J. Phys. Soc. Jpn. {\bf 71}, 491 (2002).
\bibitem{kajitani}
R. Ishikawa {\it et al.},
Jpn. J. Appl. Phys. {\bf 41}, L337, (2002). 
\bibitem{ono}
Y. Ono and T. Kajitani, {\it Oxide Thermoelectronics} (Eds. K. Kohmoto I. Terasaki, and N. Murayama, Research Signpost) 59 (2002).  
\bibitem{foot}
When we adopt a point charge model for the distorted CoO$_6$ octahedron with the parameters, i.e., 
the O-Co-O bond angle (98.5$^\circ$), 
the ratio of the Co-O bond length and Bohr radius ($\sim$4) 
and the Racah parameter ($B$ = 1065 cm$^{-1}$) for a cobalt ion 
(the value is taken from Y. Tanabe and S. Sugano J. Phys. Soc. Jpn. {\bf 9}, 766 (1954).), 
the stabilization energy of the $a_{1g}$ state against the $e'_g$ states is estimated to be 
$\sim$ 0.025 eV.  
\bibitem{sk}
J. C. Slater and G. F. Koster, Phys. Rev. {\bf 94}, 1498 (1954).  
\bibitem{koshibae1}
W. Koshibae, K. Tsutsui and S. Maekawa, Phys. Rev. B {\bf 62}, 6869 (2000). 
\bibitem{koshibae2}
W. Koshibae and S. Maekawa, Phys. Rev. Lett {\bf 87}, 236603 (2000). 
\bibitem{km}
G. Khaliullin and S. Maekawa, Phys. Rev. Lett. {\bf 85}, 3950 (2000).
\bibitem{mielke}
A. Mielke, J. Phys. A: Math. Gen. {\bf 25}, 4335 (1992).
\end{thebibliography}
\end{document}